\documentclass[epj]{webofc}
\usepackage[utf8]{inputenc}
\usepackage[varg]{txfonts}   
\usepackage{booktabs}
\usepackage{xcolor}
\definecolor{darkred}{rgb}{0.4,0.0,0.0}
\definecolor{darkgreen}{rgb}{0.0,0.4,0.0}
\definecolor{darkblue}{rgb}{0.0,0.0,0.4}
\usepackage[bookmarks,linktocpage,colorlinks,
    linkcolor = darkred,
    urlcolor  = darkblue,
    citecolor = darkgreen]{hyperref}
\usepackage{graphicx}
\usepackage{mathtools}
\usepackage{minibox}
\usepackage{tikz}
\usetikzlibrary{arrows,shapes,backgrounds,calc,positioning,patterns}
\usepackage{subfigure}
\usepackage{textpos}
\wocname{EPJ Web of Conferences}
\woctitle{Lattice2017}
\makeatletter 
\newsavebox\myboxA 
\newsavebox\myboxB 
\newlength\mylenA 

\newcommand*\xoverline[2][0.75]{%
    \sbox{\myboxA}{$\m@th#2$}%
    \setbox\myboxB\null
    \ht\myboxB=\ht\myboxA%
    \dp\myboxB=\dp\myboxA%
    \wd\myboxB=#1\wd\myboxA
    \sbox\myboxB{$\m@th\overline{\copy\myboxB}$}
    \setlength\mylenA{\the\wd\myboxA}
    \addtolength\mylenA{-\the\wd\myboxB}%
    \ifdim\wd\myboxB<\wd\myboxA%
       \rlap{\hskip 0.5\mylenA\usebox\myboxB}{\usebox\myboxA}%
    \else 
        \hskip -0.5\mylenA\rlap{\usebox\myboxA}{\hskip 0.5\mylenA\usebox\myboxB}%
    \fi}
\makeatother

\let\originalleft\left
\let\originalright\right
\renewcommand{\left}{\mathopen{}\mathclose\bgroup\originalleft}
\renewcommand{\right}{\aftergroup\egroup\originalright}
\newcommand{\e}{\operatorname{e}}

\newcommand{\of}[1]{\left(#1\right)}

\newcommand{\bof}[1]{\biggl(\bigg.#1\bigg.\biggr)}
\newcommand{\sof}[1]{\bigl(\big.#1\big.\bigr)}
\newcommand{\ssof}[1]{(#1)}
\newcommand{\fof}[1]{\left[\right.#1\left.\right]}

\newcommand{\cof}[1]{\left\{\right.#1\left.\right\}}
\newcommand{\bcof}[1]{\biggl\{\bigg.#1\bigg.\biggr\}}

\newcommand{\diag}{\operatorname{diag}}

\newcommand{\sTrace}[1]{\operatorname{Tr}\big(#1\big)}

\newcommand{\avof}[1]{\left\langle #1\right\rangle}
\newcommand{\savof}[1]{\big\langle #1\big\rangle}

\newcommand{\ket}[1]{\left| #1\right\rangle}

\newcommand{\Repart}[1]{\operatorname{Re}\left(#1\right)}
\newcommand{\Impart}[1]{\operatorname{Im}\left(#1\right)}

\newcommand{\ii}{\mathrm{i}}

\newcommand{\partd}[2]{\frac{\partial #1}{\partial #2}}

\newcommand{\spartd}[3]{\frac{\partial^{2} #1}{\partial #2 \partial #3}}

\newcommand{\order}[1]{\mathcal{O}\big(#1\big)}

\newcommand{\obs}{\mathcal{O}}

\newcommand{\abs}[1]{\left| #1\right|}

\newcommand{\op}[1]{\operatorname{#1}}
\newcommand{\umod}{\operatorname{mod}}

\renewcommand*\[{\begin{equation}}
\renewcommand*\]{\end{equation}}

\renewcommand*\hat[1]{\widehat{#1}}
\let\oldstackrel\stackrel
\renewcommand*\stackrel[2]{{\scriptstyle\oldstackrel{#1}{#2}}}

\definecolor{emphcol}{RGB}{0,0,0}
\let\oldemph\emph
\renewcommand*\emph[1]{\oldemph{\textcolor{emphcol}{#1}}}

\newlength{\hatchspread}
\newlength{\hatchthickness}
\newlength{\hatchshift}
\newcommand{\hatchcolor}{}
\tikzset{hatchspread/.code={\setlength{\hatchspread}{#1}},
         hatchthickness/.code={\setlength{\hatchthickness}{#1}},
         hatchshift/.code={\setlength{\hatchshift}{#1}},
         hatchcolor/.code={\renewcommand{\hatchcolor}{#1}}}
\tikzset{hatchspread=3pt,
         hatchthickness=0.4pt,
         hatchshift=0pt,
         hatchcolor=black}
\pgfdeclarepatternformonly[\hatchspread,\hatchthickness,\hatchshift,\hatchcolor]
   {custom north west lines}
   {\pgfqpoint{\dimexpr-2\hatchthickness}{\dimexpr-2\hatchthickness}}
   {\pgfqpoint{\dimexpr\hatchspread+2\hatchthickness}{\dimexpr\hatchspread+2\hatchthickness}}
   {\pgfqpoint{\dimexpr\hatchspread}{\dimexpr\hatchspread}}
   {
    \pgfsetlinewidth{\hatchthickness}
    \pgfpathmoveto{\pgfqpoint{\dimexpr0pt}{\dimexpr\hatchspread+\hatchshift}}
    \pgfpathlineto{\pgfqpoint{\dimexpr\hatchspread+0.15pt}{\dimexpr\hatchshift-0.15pt}}
    \ifdim \hatchshift > 0pt
      \pgfpathmoveto{\pgfqpoint{\dimexpr-0.15pt}{\dimexpr\hatchshift+0.15pt}}
      \pgfpathlineto{\pgfqpoint{\dimexpr\hatchshift+0.15pt}{\dimexpr-0.15pt}}
    \fi
    \pgfsetstrokecolor{\hatchcolor}
    \pgfusepath{stroke}
   }

\pgfdeclarepatternformonly[\hatchspread,\hatchthickness,\hatchshift,\hatchcolor]
   {custom north east lines}
   {\pgfqpoint{\dimexpr-2\hatchthickness}{\dimexpr-2\hatchthickness}}
   {\pgfqpoint{\dimexpr\hatchspread+2\hatchthickness}{\dimexpr\hatchspread+2\hatchthickness}}
   {\pgfqpoint{\dimexpr\hatchspread}{\dimexpr\hatchspread}}
   {
    \pgfsetlinewidth{\hatchthickness}
    \pgfpathmoveto{\pgfqpoint{0pt}{\dimexpr\hatchshift}}
    \pgfpathlineto{\pgfqpoint{\dimexpr\hatchspread+0.15pt}{\dimexpr\hatchspread+0.15pt+\hatchshift}}
    \ifdim \hatchshift > 0pt
      \pgfpathmoveto{\pgfqpoint{\dimexpr\hatchspread-\hatchshift-0.15pt}{\dimexpr-0.15pt}}
      \pgfpathlineto{\pgfqpoint{\dimexpr\hatchspread+0.15pt}{\dimexpr\hatchshift+0.15pt}}
    \fi
    \pgfsetstrokecolor{\hatchcolor}
    \pgfusepath{stroke}
   }

\pgfdeclarepatternformonly[\hatchspread,\hatchthickness,\hatchshift,\hatchcolor]
   {custom vertical lines}
   {\pgfqpoint{\dimexpr-2\hatchthickness}{\dimexpr-2\hatchthickness}}
   {\pgfqpoint{\dimexpr\hatchspread+2\hatchthickness}{\dimexpr\hatchspread+2\hatchthickness}}
   {\pgfqpoint{\dimexpr\hatchspread}{\dimexpr\hatchspread}}
   {
    \pgfsetlinewidth{\hatchthickness}
    \pgfpathmoveto{\pgfqpoint{\dimexpr\hatchshift}{0pt}}
    \pgfpathlineto{\pgfqpoint{\dimexpr\hatchshift}{\dimexpr\hatchspread+0.15pt}}
    \pgfsetstrokecolor{\hatchcolor}
    \pgfusepath{stroke}
   }

\pgfdeclarepatternformonly[\hatchspread,\hatchthickness,\hatchshift,\hatchcolor]
   {custom horizontal lines}
   {\pgfqpoint{\dimexpr-2\hatchthickness}{\dimexpr-2\hatchthickness}}
   {\pgfqpoint{\dimexpr\hatchspread+2\hatchthickness}{\dimexpr\hatchspread+2\hatchthickness}}
   {\pgfqpoint{\dimexpr\hatchspread}{\dimexpr\hatchspread}}
   {
    \pgfsetlinewidth{\hatchthickness}
    \pgfpathmoveto{\pgfqpoint{0pt}{\dimexpr\hatchshift}}
    \pgfpathlineto{\pgfqpoint{\dimexpr\hatchspread+0.15pt}{\dimexpr\hatchshift}}
    \pgfsetstrokecolor{\hatchcolor}
    \pgfusepath{stroke}
   }

\tikzset{cross/.style={cross out,draw,minimum size=2*(#1-\pgflinewidth),inner sep=0pt, outer sep=0pt}}

\begin{document}
%
\selectlanguage{english}
\title{%
\begin{textblock*}{100pt}(337pt,-116pt)
\textnormal{\small \texttt{HIP-2017-31/TH}}
\end{textblock*}Spin models in complex magnetic fields: a hard sign problem
}
\author{%
\firstname{Philippe} \lastname{de Forcrand}\inst{1,2}
\and
\firstname{Tobias} \lastname{Rindlisbacher}\inst{3}\fnsep\thanks{Speaker, \email{tobias.rindlisbacher@helsinki.fi}}
}
\institute{%
ETH Z\"urich, Institute for Theoretical Physics, Wolfgang-Pauli-Str. 27, CH-8093 Z\"urich, Switzerland
\and
CERN, Theory Division, CH-1211 Geneva 23, Switzerland
\and
University of Helsinki, Department of Physics, P.O. Box 64, FI-00014 University of Helsinki, Finland
}
\abstract{%
Coupling spin models to complex external fields can give rise to interesting phenomena like zeroes of the partition function (Lee-Yang zeroes, edge singularities) or oscillating propagators. Unfortunately, it usually also leads to a severe sign problem that can be overcome only in special cases; if the partition function has zeroes, the sign problem is even representation-independent at these points. In this study, we couple the N-state Potts model in different ways to a complex external magnetic field and discuss the above mentioned phenomena and their relations based on analytic calculations (1D) and results obtained using a modified cluster algorithm (general D) that in many cases either cures or at least drastically reduces the sign-problem induced by the complex external field.
}
\maketitle
\section{Introduction}\label{intro}

\subsection{Parameters of the Potts partition function}\label{ssec:pottsparam}
A general partition function for a $d$-dimensional $N$-state Potts system of size $V$, in which the spins couple linearly to a complex external field, can be written as:
\[
Z_{N,V}\of{\beta,\,h_{0},\,\ldots,\,h_{N-1}}\,=\,\sum\limits_{\cof{s}}\,\exp\bof{\sum\limits_{x}\bof{\beta\,\sum\limits_{\nu=1}^{d}\of{2\,\delta_{s_{x},s_{x+\hat{\nu}}}-1}\,+\,\sum\limits_{n=0}^{N-1}\,h_{n}\,\delta_{n,s_{x}}}}\ ,\label{eq:pottspartfgen}
\]
where $s_{x}\in\cof{0,\ldots,N-1}$ is the Potts spin on site $x$, $\beta\,\in\,\mathbb{R}_{+}$ is the inverse temperature, and $h_{n}\,\in\,\mathbb{C}$ , $n\,\in\,\cof{0,\ldots,N-1}$ are $N$ complex parameters that define the coupling to the external fields (note that only $\of{N-1}$ of the $h_{n}$ are linearly independent as $\sum_{n=0}^{N-1}\,\delta_{n,s_{x}}=1$ $\forall x$). We will focus on a subset of possible choices for the external fields, namely on the cases where $h_{n}=h\,\e^{\frac{2\,\pi\,\ii\,n}{N}}+h'\e^{-\frac{2\,\pi\,\ii\,n}{N}}$, with $h,h'\in\mathbb{C}$, so that the partition function can be written as:
\[
Z_{N,V}\of{\beta,\,h,\,h'}\,=\,\sum\limits_{\cof{s}}\,\exp\bof{\sum\limits_{x}\bof{\beta\,\sum\limits_{\nu=1}^{d}\of{2\,\delta_{s_{x},s_{x+\hat{\nu}}}-1}\,+\,h\,P_{x}\,+\,h'\,P^{*}_{x}}}\ ,\label{eq:pottspartf0}
\]
with $P_{x}=\e^{\frac{2\,\pi\,\ii\,s_{x}}{N}}$ and $P^{*}_{x}=\e^{-\frac{2\,\pi\,\ii\,s_{x}}{N}}$. Observables of interest will be the magnetizations,
\[
\avof{P}\,=\,\frac{1}{V}\partd{\log\of{Z_{N,V}\of{\beta,\,h,\,h'}}}{h}\quad\text{and}\quad \avof{P^{*}}\,=\,\frac{1}{V}\partd{\log\of{Z_{N,V}\of{\beta,\,h,\,h'}}}{h'}\ ,\label{eq:magnetizations}
\]
and the two-point function
\[
\avof{P_{x}\,P^{*}_{y}}_{c}\,=\,\frac{1}{V}\spartd{\log\of{Z_{N,V}\of{\beta,\,h,\,h'}}}{h_{x}}{h'_{y}}\bigg|_{h_{z}=h,h'_{z}=h'\forall z}\ ,\label{eq:twopointfunc}
\]
where $h$ and $h'$ are temporarily interpreted as site-dependent quantities.  

\subsection{Sign problem}\label{ssec:signproblem}
For arbitrary $h,\,h'\in\mathbb{C}$, the configuration weight (i.e. the exponential) in the partition sum \eqref{eq:pottspartf0} has in general a non-constant complex phase and lacks therefore a direct probabilistic interpretation, which would be necessary to estimate observables using Monte Carlo. This is called a \emph{sign-problem} and a standard way to deal with it is by using \emph{reweighting} techniques, which means that one does Monte Carlo for the corresponding \emph{phase-quenched} system:
\[
Z_{N,V,q}\of{\beta,h,h'}=\sum\limits_{\cof{s}}\,\abs{w\fof{s}}\ ,\quad\text{where}\quad w\fof{s}=\exp\bof{\sum\limits_{x}\bof{\beta\,\sum\limits_{\nu=1}^{d}\of{2\,\delta_{s_{x},s_{x+\hat{\nu}}}-1}\,+\,h\,P_{x}\,+\,h'\,P^{*}_{x}}}\ , 
\]
and evaluates observables for the original system by using that
\[
\avof{\obs\fof{s}}\,=\,\frac{\sum\limits_{\cof{s}}\,\obs\fof{s}\,w\fof{s}}{\sum\limits_{\cof{s}}\,w\fof{s}}\,=\,\frac{\sum\limits_{\cof{s}}\,\obs\fof{s}\,R\fof{s}\,\abs{w\fof{s}}}{\sum\limits_{\cof{s}}\,R\fof{s}\abs{w\fof{s}}}\,=\,\frac{\avof{\obs\fof{s}\,R\fof{s}}_{q}}{\avof{R\fof{s}}_{q}}\ ,\label{eq:obsreweighting}
\]
where $R\fof{s}=w\fof{s}/\abs{w\fof{s}}$ is the complex phase of $w\fof{s}$ and $\avof{\obs'\fof{s}}_{q}$ refers to the expectation value of an observable $\obs'\fof{s}$ with respect to the phase-quenched system, i.e.:
\[
\avof{\obs'\fof{s}}_{q}\,=\,\frac{1}{Z_{N,V,q}\of{\beta,h,h'}}\,\sum\limits_{\cof{s}}\,\obs'\fof{s}\cdot\abs{w\fof{s}}\ .
\]
The reweighting formula \eqref{eq:obsreweighting} is a priori exact, but if the expectation values $\avof{\ldots}_{q}$ after the last equality sign in \eqref{eq:obsreweighting} are evaluated by Monte Carlo, then the statistical fluctuations of the phase $R\fof{s}$ cause problems: by writing  
\[
\avof{R\fof{s}}_{q}\,=\,\frac{Z_{N,V}\of{\beta,h,h'}}{Z_{N,V,q}\of{\beta,h,h'}}\,=\,\e^{-V\,\Delta f}\ ,\label{eq:avsignfreenergy}
\]
where $\Delta f=f-f_{q}$ is the difference between the free energy densities for the full and the phase-quenched system (which becomes independent of the system size $V$ if $V$ is large), we see that (the modulus of) $\avof{R\fof{s}}$ decays exponentially with increasing $V$, which in turn implies exponential growth of the fluctuations in the reweighted observable $\obs\fof{s}\,R\fof{s}/\avof{R\fof{s}}_{q}$ and the corresponding statistical error. As in a Monte Carlo simulation the statistical error depends on the number $N_{samp}$ of measurements like $\text{err.} \propto 1/\sqrt{N_{samp}}$, the number of measurements required in \eqref{eq:obsreweighting} to obtain equally accurate estimates of $\avof{\obs\fof{s}}$ for different system sizes $V$, would therefore scale like $N_{samp}\,\propto\,\e^{2\,V\,\Delta f}$, which limits the applicability of the reweighting method \eqref{eq:obsreweighting} to very small system sizes.

\section{Exact solution in 1D}\label{sec:exactsol}
In 1D, the partition function \eqref{eq:pottspartfgen} can be computed analytically in terms of the eigenvalues of the transfer matrix:
\[
Z_{N,L}\of{\beta,\,h,\,h'}\,=\,\sTrace{T^{L}\of{\beta,h,h'}}\,=\,\sTrace{\Lambda^{L}\of{\beta,h,h'}}\,=\,\sum\limits_{k=1}^{N}\,\lambda_{k}^{L}\of{\beta,h,h'}\ ,\label{eq:partf1d}
\]
where $L$ is the system size, $T\of{\beta,h,h'}$ the transfer-matrix,
\[
T_{s_{x},s_{x+1}}\of{\beta,h,h'}\,=\,\exp\of{\beta\,\sof{2\,\delta_{s_{x},s_{x+1}}\,-\,1}\,+\,h\,\frac{P_{x}\,+\,P_{x+1}}{2}\,+\,h'\,\frac{P^{*}_{x}\,+\,P^{*}_{x+1}}{2}}\ ,
\]
and $\Lambda\of{\beta,h,h'}=\diag\cof{\lambda_{1}\of{\beta,h,h'},\ldots,\lambda_{N}\of{\beta,h,h'}}$ is the diagonal matrix of eigenvalues of $T\of{\beta,h,h'}$, so that:
\[
T\of{\beta,h,h'}\,=\,U^{-1}\of{\beta,h,h'}\,\Lambda\of{\beta,h,h'}\,U\of{\beta,h,h'}\quad\text{with}\quad U\of{\beta,h,h'}\,\in\,\op{L}\of{N,\mathbb{C}}\ .
\]

\subsection{Non-Hermitian transfer matrix}\label{ssec:nonhermtransmatrix}
\begin{figure}[!h]
\centering
\begin{minipage}[b]{0.49\linewidth}
\centering
\begin{tikzpicture}[scale=0.92, every node/.style={transform shape}]
  \node[inner sep=0pt,opacity=1]{\includegraphics[width=\linewidth]{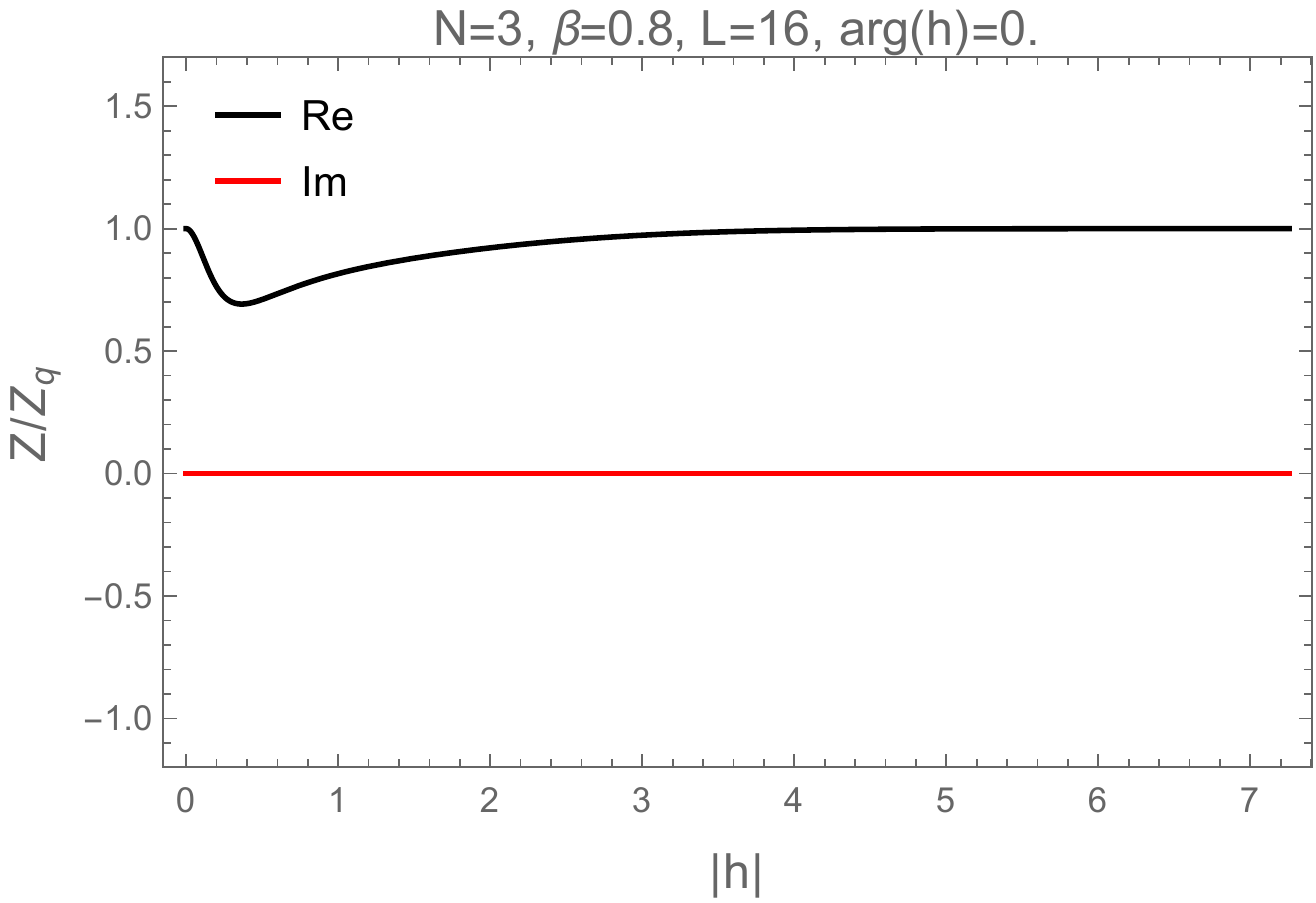}};
  \node[fill=blue!20,inner sep=3pt] at (13pt,-35pt){$Z\,\in\,\mathbb{R}_{+}$};
\end{tikzpicture}
\end{minipage}\hfill
\begin{minipage}[b]{0.45\linewidth}
\centering
\begin{tikzpicture}[scale=0.88, every node/.style={transform shape}]
  \node[draw,circle,inner sep=0pt,color=red] (C) at (86.5pt,81.0pt) {};
  \node[inner sep=0pt,above right,opacity=1]{\includegraphics[width=\textwidth]{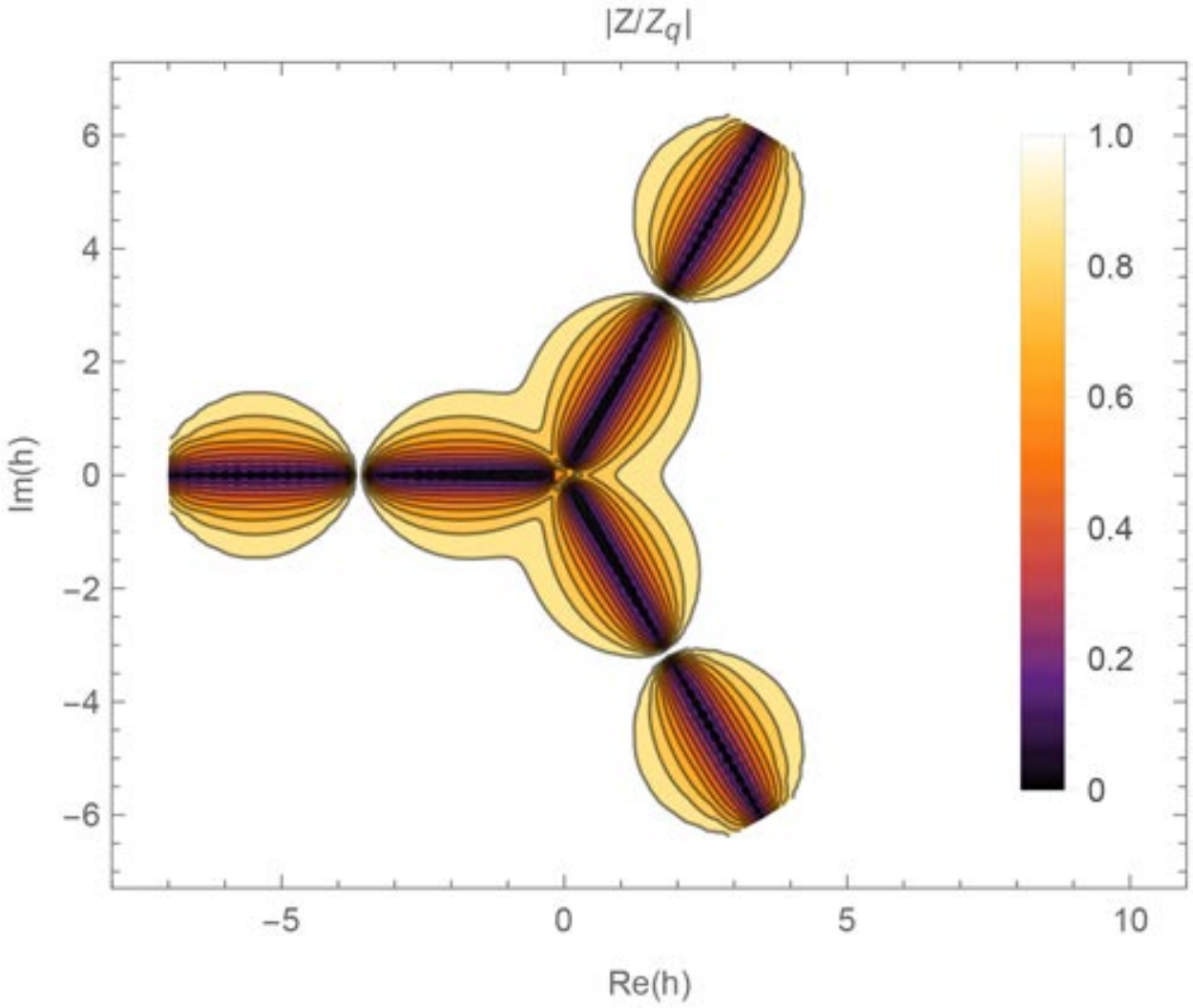}};
  \draw[->,red,thick] (C) -- +(0:45pt);
\end{tikzpicture}
\end{minipage}\\
\begin{minipage}[b]{0.49\linewidth}
\centering
\begin{tikzpicture}[scale=0.92, every node/.style={transform shape}]
  \node[inner sep=0pt,opacity=1]{\includegraphics[width=\linewidth]{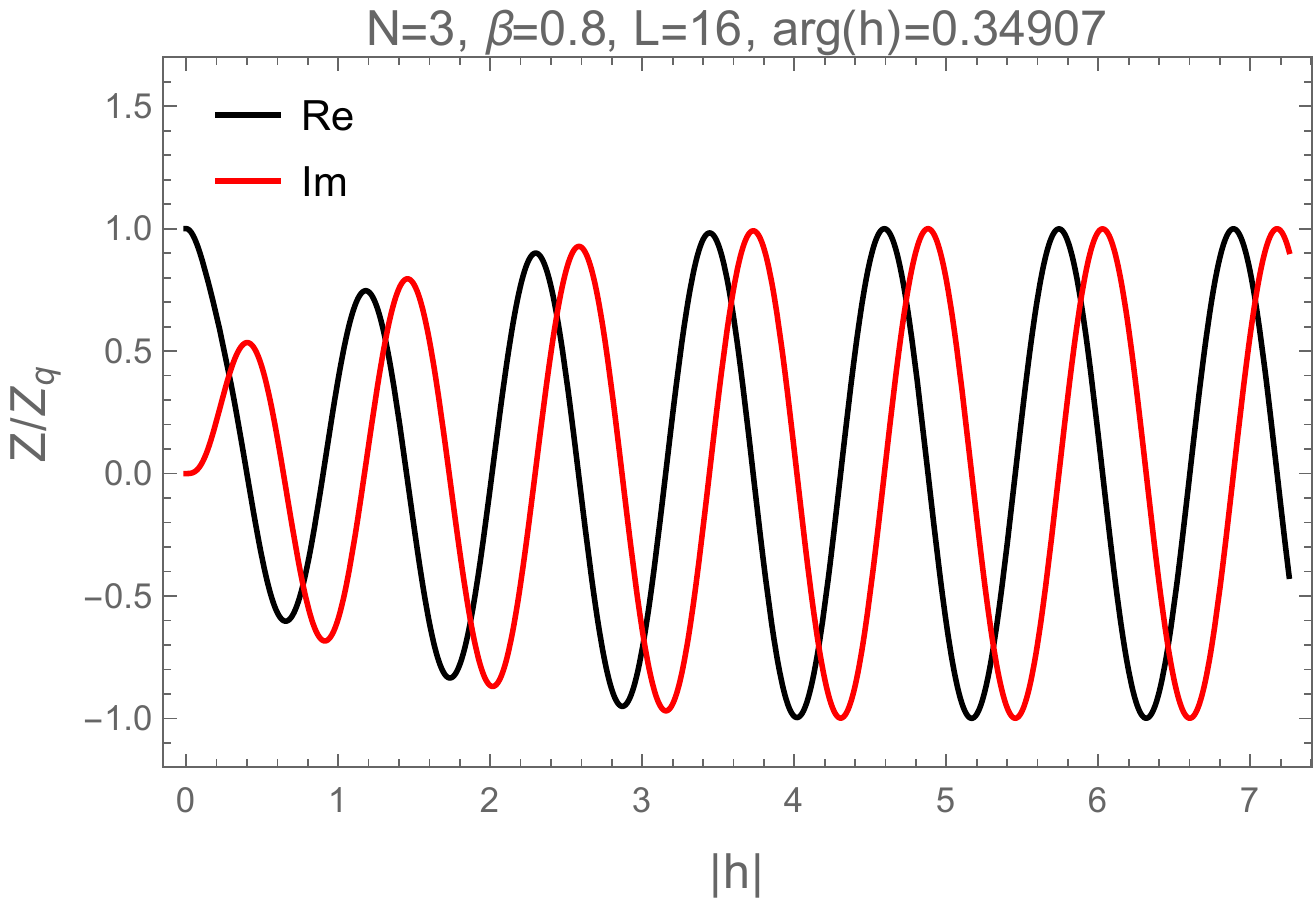}};
  \node[fill=red!20,inner sep=3pt] at (13pt,-35pt){$Z\,\in\,\mathbb{C}$};
\end{tikzpicture}
\end{minipage}\hfill
\begin{minipage}[b]{0.45\linewidth}
\centering
\begin{tikzpicture}[scale=0.88, every node/.style={transform shape}]
  \node[draw,circle,inner sep=0pt,color=red] (C) at (86.5pt,81.0pt) {};
  \node[inner sep=0pt,above right,opacity=1]{\includegraphics[width=\textwidth]{img/potts_1d_absz_vs_reh_and_imh_l16_n3_b08.pdf}};
  \def\angl{180/3.14159*0.34907};
  \draw[->,red,thick] (C) -- +(\angl:45pt);
\end{tikzpicture}
\end{minipage}\\
\begin{minipage}[b]{0.49\linewidth}
\centering
\begin{tikzpicture}[scale=0.92, every node/.style={transform shape}]
  \node[inner sep=0pt,opacity=1]{\includegraphics[width=\textwidth]{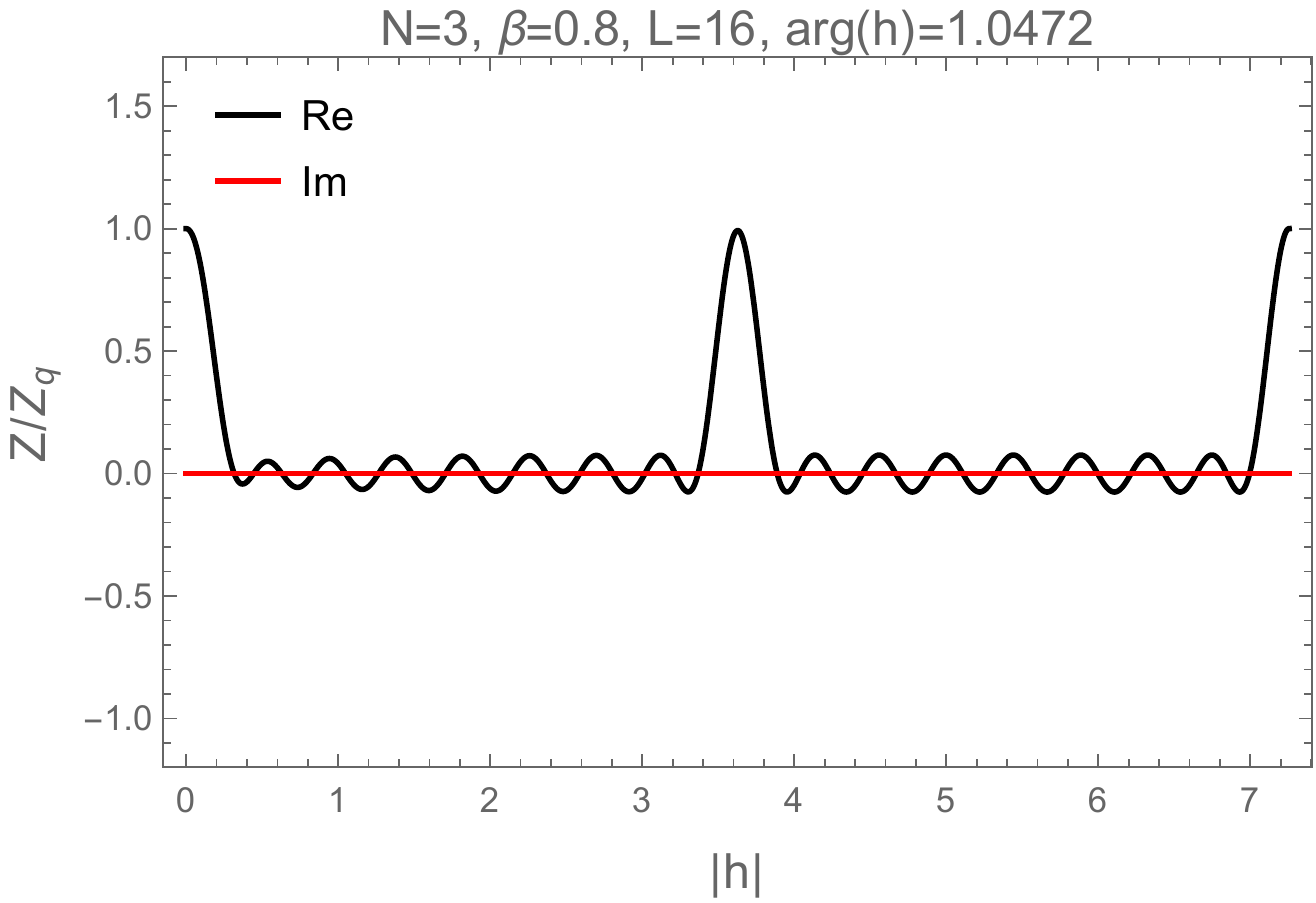}};
    \node[pattern=custom horizontal lines,hatchspread=4pt,hatchthickness=1pt,hatchshift=2pt,hatchcolor=blue!20,inner sep=3pt] at (13pt,-35pt){\phantom{$Z\,\in\,\mathbb{R}$}};
    \node[pattern=custom horizontal lines,hatchspread=4pt,hatchthickness=1pt,hatchshift=0pt,hatchcolor=red!20,inner sep=3pt] at (13pt,-35pt){$Z\,\in\,\mathbb{R}$};
    \node[draw,rectangle,thick,color=red,fill=white,inner sep=1pt] at (12pt,-18pt){\minibox{ $Z=0$ $\Rightarrow$ irreducible sign-problem! }};
\end{tikzpicture}
\end{minipage}\hfill
\begin{minipage}[b]{0.45\linewidth}
\centering
\begin{tikzpicture}[scale=0.88, every node/.style={transform shape}]
  \node[draw,circle,inner sep=0pt,color=red] (C) at (86.5pt,81.0pt) {};
  \node[inner sep=0pt,above right,opacity=1]{\includegraphics[width=\textwidth]{img/potts_1d_absz_vs_reh_and_imh_l16_n3_b08.pdf}};
  \def\angl{180/3.14159*1.0472};
  \draw[->,red,thick] (C) -- +(\angl:45pt);
\end{tikzpicture}
\end{minipage}
\caption{The figures illustrate how for a 1d system defined by \eqref{eq:partf1d} with $N=3$, $L=16$, $\beta=0.8$ and $h'=0$, the quantity $Z_{N,L}\of{\beta,h,0}/Z_{N,L,q}\of{\beta,h,0}$ (which is a measure for the severity of the sign problem, see eq.~\eqref{eq:avsignfreenergy}) depends on the direction and magnitude of $h\in\mathbb{C}$. For $\arg\of{h}=0$ (top row), $Z_{N,L}$ is real and positive, and the sign-problem can most likely be overcome by a clever choice of new configuration space coordinates. For $0<\arg\of{h}<\pi/N$ (middle row), $Z_{N,L}$ is complex but the sign-problem might still be overcome by a clever change of new coordinates in configuration space (as the complex phase of $Z_{N,L}$ could in some coordinates just be a constant for all configurations). Finally, for $\arg\of{h}=\pi/N$ (bottom row), $Z_{N,L}$ is again real but no longer positive-definite. At the zeroes of $Z_{N,L}$, the sign-problem is irreducible and will be present in all possible representations of the partition function. Note that $Z_{N,L}\of{\beta,h,0}$ is invariant under $Z_{N}$ rotations of the complex $h$.}
\label{fig:signproblemvsh1d}
\end{figure}

If the transfer matrix $T\of{\beta,h,h'}$ is non-Hermitian, then its eigenvalues $\lambda_{k}$ are in general complex, and therefore also the partition function \eqref{eq:partf1d} is in general complex (see Fig.~\ref{fig:signproblemvsh1d}, middle row). However, as shown in \cite{Meisinger:2010be}, if the Hamiltonian of the theory under consideration is invariant under $\mathcal{CT}$-transformations, then we have that
\[
H\,\ket{\psi}\,=\,E\,\ket{\psi}\,\Rightarrow\,H\,\mathcal{C}\mathcal{T}\,\ket{\psi}\,=\,\mathcal{C}\mathcal{T}\,H\,\ket{\psi}\,=\,\mathcal{C}\mathcal{T}\,E\,\ket{\psi}\,=\,E^{*}\,\mathcal{C}\mathcal{T}\,\ket{\psi}\ ,
\]
so that the energies are either real or appear in complex-conjugate pairs. This implies that also the eigenvalues of the transfer matrix have either to be real or appear in complex-conjugate pairs and \eqref{eq:partf1d} is therefore not complex but real as well! In our case, the Hamiltonian is given by:
\[
H=-\sum\limits_{x}\bof{\of{2\,\delta_{s_{x},s_{x+1}}-1}\,+\,\frac{h\,P_{x}\,+\,h'\,P^{*}_{x}}{\beta}}\ ,\label{eq:pottshamiltonian}
\]
which has $\mathcal{CT}$-symmetry if one sets for example $h+h'=h_{R}$, $h-h'=h_{I}$ with $h_{R},\,h_{I}\in\mathbb{R}$ (see \cite{Meisinger:2010be,Akerlund:2016myr}), or if we set $h'=0$ and $h\in\mathbb{C}$ with $\arg\of{h}\in\cof{\frac{\pi\,k}{N}\,|\,k\in\mathbb{Z}}$ (see Fig.~\ref{fig:signproblemvsh1d}, first and last row) and use $\mathcal{SCT}$ instead of $\mathcal{CT}$ with $\mathcal{S}\in S_{N}$ (permutation group). 

\subsection{Edge singularities, disorder lines and two-point functions}\label{ssec:edgesinganddolines}
Assume that all eigenvalues of the transfer matrix are either real or part of a complex-conjugate pair, i.e. $\forall n\,\in\,\cof{1,\ldots,N} \exists m\,\in\cof{1,\ldots,N}\,:\,\lambda_{m}=\lambda^{*}_{n}$, and that they are ordered according to their moduli: $\abs{\lambda_{1}}\geq\abs{\lambda_{2}}\geq\ldots\geq\abs{\lambda_{N}}\,|\,\lambda_{n}\neq\lambda_{m}\,\forall\,m,n\,\in\,\cof{1,\ldots,N}$. It is then possible, to identify three prototypes of "phases" \cite{Meisinger:2010be,Akerlund:2016myr}, in which the two-point function,
\[
\savof{P_{0}\,P^{*}_{x}}\,=\,\frac{\sTrace{\phi\,T^{x}\,\phi^{*}\,T^{L-x}}}{\sTrace{T^{L}}}\,=\,\frac{\sTrace{\tilde{\phi}\,\Lambda^{x}\,\tilde{\phi}^{*}\,\Lambda^{L-x}}}{\sTrace{\Lambda^{L}}}\ ,\label{eq:twopointfunc1d}
\]
with $\phi_{k l}\,=\,\delta_{k l}\,\e^{\frac{2\,\pi\,\ii\,\of{k-1}}{N}}$ , $\phi^{*}_{k l}\,=\,\delta_{k l}\,\e^{-\frac{2\,\pi\,\ii\,\of{k-1}}{N}}$, $\tilde{\phi}\,=\,U\,\phi\,U^{-1}$ , $\tilde{\phi}^{*}\,=\,U\,\phi^{*}\,U^{-1}$, behaves very differently.\\
In the limit $L\to\infty$, the origin of these different phases can be understood by analyzing the  dependency of \eqref{eq:twopointfunc1d} on the transfer-matrix eigenvalues $\lambda_{n},\,n\in\cof{1,\ldots,N}$ \cite{Akerlund:2016myr}:
\begin{enumerate}
\item if $\lambda_{n}\in\mathbb{R}_{+}\,\forall\,n\in\cof{1,\ldots,N}$, we are in the "gaseous" phase, where $Z_{N,L}\approx\lambda_{1}^{L}\in\mathbb{R}_{+}$ and the two-point function is a pure sum of exponentials:
\[
\savof{P_{0}\,P^{*}_{x}}=\sum\limits_{n=1}^{N}\,\abs{\tilde{\phi}_{1 n}}^{2}\,\of{\frac{\lambda_{n}}{\lambda_{1}}}^{x}\,=\,\abs{\tilde{\phi}_{1 1}}^{2}+\abs{\tilde{\phi}_{1 2}}^{2}\,\e^{-m_{1}\,x}+\order{\e^{-m_{2}\,x}}\quad\text{with}\quad \e^{-m_{n}}=\frac{\lambda_{n}}{\lambda_{1}}\ .
\]
\item If $\lambda_{1}\in\mathbb{R}_{+}$ and $\lambda_{2}=\lambda_{3}^{*}\in\mathbb{C}$, we are in the "liquid" phase where still $Z_{N,L}\approx\lambda_{1}^{L}\in\mathbb{R}_{+}$ but where the two-point function is now a damped oscillation:
\[
\savof{P_{0}\,P^{*}_{x}}\approx \abs{\tilde{\phi}_{0 0}}^{2}\,+\,\abs{\tilde{\phi}_{0 1}}^{2}\,\of{\of{\frac{\lambda_{2}}{\lambda_{1}}}^{x}+\of{\frac{\vphantom{\lambda_{2}}\smash{\lambda_{2}^{*}}}{\lambda_{1}}}^{x}}\,=\,\abs{\tilde{\phi}_{0 0}}^{2}\,+\,2\,\abs{\tilde{\phi}_{0 1}}^{2}\,\e^{-m_{R}\,x}\,\cos\of{m_{I}\,x}\ ,
\]
where the real and imaginary masses $m_{R}$ and $m_{I}$ are given by the relation $\e^{-m_{R}+\ii\,m_{I}}=\frac{\lambda_{2}}{\lambda_{1}}$ .
\item Finally, if $\lambda_{1}=\lambda_{2}^{*} \in \mathbb{C}$, we are in the crystalline phase, where $Z_{N,L}\approx\lambda_{1}^{L}+\ssof{\lambda_{1}^{*}}^{L}=2\,\abs{\lambda_{1}}^{L}\,\cos\of{m_{I}\,L/2}$ and the two-point function is a pure oscillation:
\[
\savof{P_{0}\,P^{*}_{x}}\approx \frac{\lambda_{1}^{L-x}\sum\limits_{n=1}^{N}\,\abs{\tilde{\phi}_{1 n}}^{2}\,\lambda_{n}^{x}\,+\,\ssof{\lambda_{1}^{*}}^{L-x}\sum\limits_{n=1}^{N}\,\abs{\tilde{\phi}_{2 n}}^{2}\,\lambda_{n}^{x}}{\lambda_{1}^{L}+\ssof{\lambda_{1}^{*}}^{L}}\approx \abs{\tilde{\phi}_{1 1}}^{2}\,+\,\frac{\cos\of{m_{I}\,\of{x-L/2}}}{\cos\of{m_{I}\,L/2}}\,\abs{\tilde{\phi}_{1 2}}^{2}\ ,
\]
where the mass has now only an imaginary part $m_{I}$, defined through $\e^{\ii\,m_{I}}=\frac{\lambda_{2}}{\lambda_{1}}=\frac{\lambda^{*}_{1}}{\lambda_{1}}$ .
\end{enumerate}
Strictly speaking, the gaseous and liquid phases are not really distinct phases as they are not separated by a true phase transition but only by a so-called \emph{disorder line}. The boundary between gaseous or liquid phase and the crystalline phase on the other hand, forms an \emph{edge-singularity} where $Z_{N,L}\of{\beta,h,h'}=0$ and the free energy is therefore discontinuous along such a boundary.\\
A zero of the partition function implies that there must be a sign-problem which is physical or \emph{irreducible}, in the sense, that it cannot possibly be removed by a change or representation for the partition sum: if the partition sum is zero, this means that either all configurations that contribute to it have weight zero, or that there must be cancellations between weights for different configurations, which means that some of them must be negative. In Sec.~\ref{sec:results}, we will investigate how quickly the average sign drops to zero when approaching an edge-singularity in different representations of the same partition function.  


\section{Alternative representations}\label{sec:altrep}
From the discussion in Sec.~\ref{ssec:signproblem}, it is clear that the severity of the sign-problem in the partition function \eqref{eq:pottspartf0} depends on the choice of representation (spin, flux-variable, cluster, etc.?). It is therefore reasonable, to investigate how well the different representations deal with the sign-problem in the different cases illustrated in Fig.~\ref{fig:signproblemvsh1d}.

\subsection{Flux-variable representation}\label{ssec:fluxrep}
The flux-variable representation for \eqref{eq:pottspartf0} is unfortunately only available for $N=2,3$ in which case the $N$-state Potts model is equivalent to the so-called $N$-state \emph{clock model}. For $N=2$, the flux-variable formulation of \eqref{eq:pottspartf0} can be written as \cite{Wolff:2008km}:
\begin{multline}
Z\of{\beta,h,h'}\,=\,\of{\cosh\of{h+h'}\,\cosh\of{\beta}^d}^{V}\\
\cdot\sum\limits_{\cof{k,m}}\,\prod\limits_{x}\bcof{\tanh\of{h+h'}^{m_{x}}\,\tanh\of{\beta}^{\sum_{\mu=1}^{d}\,k_{x,\mu}}\,\delta_{0,\umod_{2}\sof{m_{x}+\sum_{\mu=1}^{d}\sof{k_{x,\mu}+k_{x-\hat{\mu},\mu}}}}} .\label{eq:fluxpartf0}
\end{multline}
with the flux variables $k_{x,\mu}\in\cof{0,1}$ living on links, and the monomer occupation numbers $m_{x}\in\cof{0,1}$ living on sites. The configurations weights are obviously complex for general $h,h'\in\mathbb{C}$, real for $h+h'$ purely real or purely imaginary, and real and positive for $h+h'$ real and positive. For the flux-variable formulation of the $N=3$ case, see \cite{Mercado:2012yf}.

\subsection{Cluster representation}\label{ssec:clusterrep}
Deriving the cluster representation for the Potts partition function \eqref{eq:pottspartfgen} is straightforward \cite{Swendsen:1987ce}. One can then make use of the fact that individual clusters are completely decoupled so that one can easily sum over all possible spin states in each of them \cite{Alford:2001ug}. The resulting partition function reads
\[
Z_{N}\of{\beta,\,h_{0},\,\ldots,\,h_{N-1}}\,=\,\e^{-\beta\,V\,d}\sum\limits_{\cof{b}}\,\sof{\e^{2\,\beta}-1}^{\sum\limits_{x,\nu}b_{x,\nu}}\,\prod\limits_{\mathclap{C\in C\fof{b}}}\,W_{N,h_{0},\ldots,h_{N-1}}\of{\abs{C}}\ ,\label{eq:clusterpartfgen}
\]
with $b_{x,\nu}$ being the bond-variable that connects site $x$ with site $x+\hat{\nu}$. Sites that are connected by a bond correspond to the same cluster, so that each bond-configuration $b=\cof{b_{x,\nu}}_{x,\nu}$ defines a corresponding set of clusters $C\fof{b}$. The cluster weights are given in terms of a sum over all possible cluster spins:
\[
W_{N,h_{0},\ldots,h_{N-1}}\of{V_{C}}\,=\,\sum\limits_{s=0}^{N-1}\exp\bof{V_{C}\,\sum\limits_{n=0}^{N-1}\,h_{n}\,\delta_{n,s}}\,=\,\sum\limits_{n=0}^{N-1}\exp\of{V_{C}\,h_{n}}\ .\label{eq:clusterweight}
\]
If \eqref{eq:clusterweight} is real and positive for all $V_{C}\in\cof{1,\ldots,V}$, then the representation \eqref{eq:clusterpartfgen} is obviously sign-problem free. By writing \eqref{eq:clusterweight} as:
\[
W_{N,h_{0},\ldots,h_{N-1}}\of{\abs{C}}\,=\,{\sum\limits_{n=0}^{N-1}\exp\of{\abs{C}\,\Repart{h_{n}}}\,\cos\of{\abs{C}\,\Impart{h_{n}}}}\,+\,{\ii\,\sum\limits_{n=0}^{N-1}\exp\of{\abs{C}\,\Repart{h_{n}}}\,\sin\of{\abs{C}\,\Impart{h_{n}}}}\label{eq:clusterweightrw}
\]
we see from the second sum in \eqref{eq:clusterweightrw} that $W_{N,h_{0},\ldots,h_{N-1}}\of{V_{C}}\,\in\,\mathbb{R}\ \forall\,V_{C}\,\in\,\cof{1,\ldots,V}$ if non-real $h_{n}$ occur in complex-conjugate pairs, and that $W_{N,h_{0},\ldots,h_{N-1}}\of{V_{C}}\,\in\,\mathbb{R}_{+}\ \forall\,V_{C}\,\in\,\cof{1,\ldots,V}$ if there is in addition at least one purely real $h_{n}$ that is sufficiently large so that the first sum in \eqref{eq:clusterweightrw} is always positive.\\
If we set $h_{n}=h\,\e^{\frac{2\,\pi\,\ii\,n}{N}}=\abs{h}\,\e^{\ii\,\arg\of{h}\frac{2\,\pi\,\ii\,n}{N}}$, which corresponds to setting $h'=0$ and $h\in\mathbb{C}$  in \eqref{eq:pottspartf0}, the condition that non-real $h_{n}$ appear in complex-conjugate pairs is satisfied if $\arg{h}=\frac{\pi\,k}{N}$ with $k\in\mathbb{Z}$. If $k$ is even, then the $h_{n}$ which has the largest real part is purely real and $W_{N,h}\of{V_{C}}\in\mathbb{R}_{+}$, whereas if $k$ is odd, the $h_{n}$ with the largest real part form a complex-conjugate pair and $W_{N,h}\of{V_{C}}\in\mathbb{R}$, i.e. can be negative.

\section{Results}\label{sec:results}

\subsection{Severity of sign-problem in different representations}\label{ssec:severityofsignproblem}
\begin{figure}[h]
\centering
\begin{minipage}[t]{0.45\linewidth}
\centering
$\beta=0.3$, $N=2$\\[-5pt]
\includegraphics[width=\linewidth]{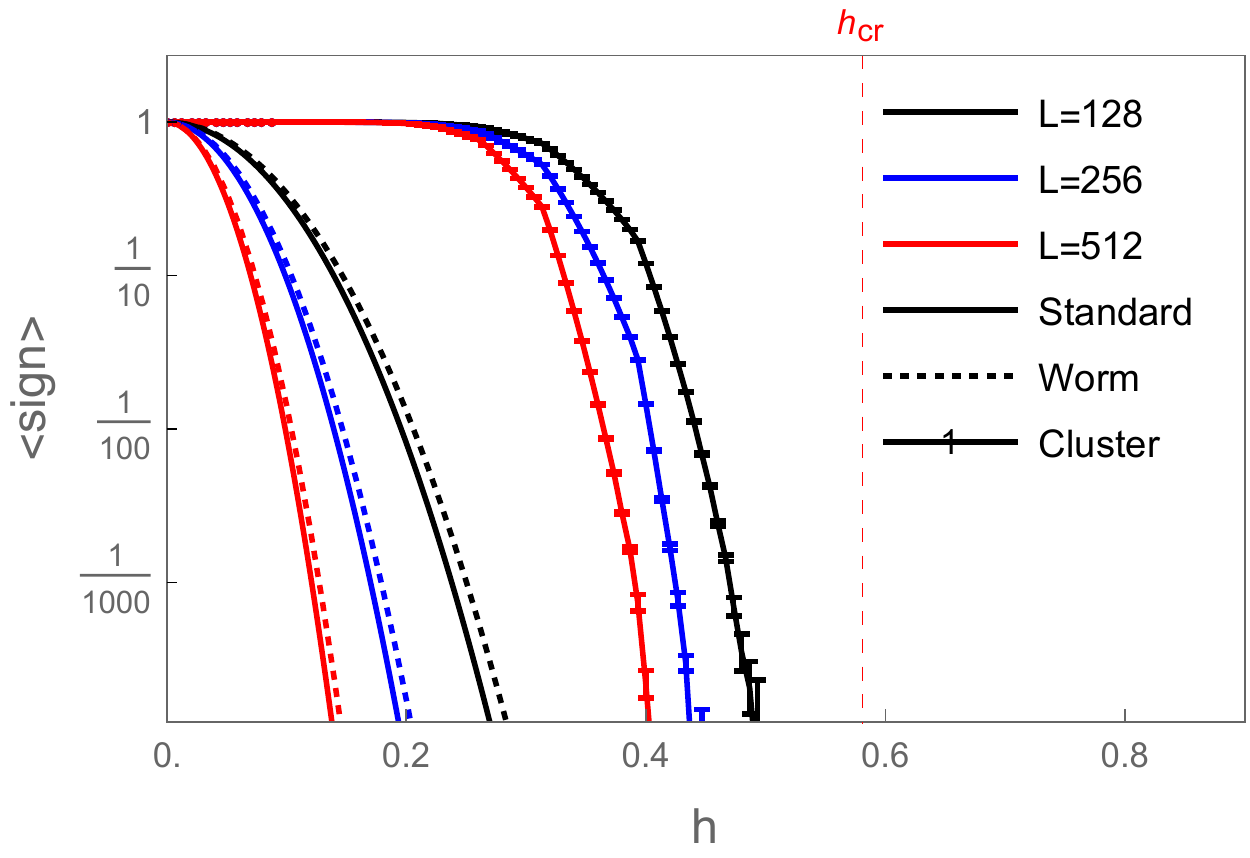}
\end{minipage}\hfill
\begin{minipage}[t]{0.45\linewidth}
\centering
$\beta=0.5$, $N=3$\\[-5pt]
\includegraphics[width=\linewidth]{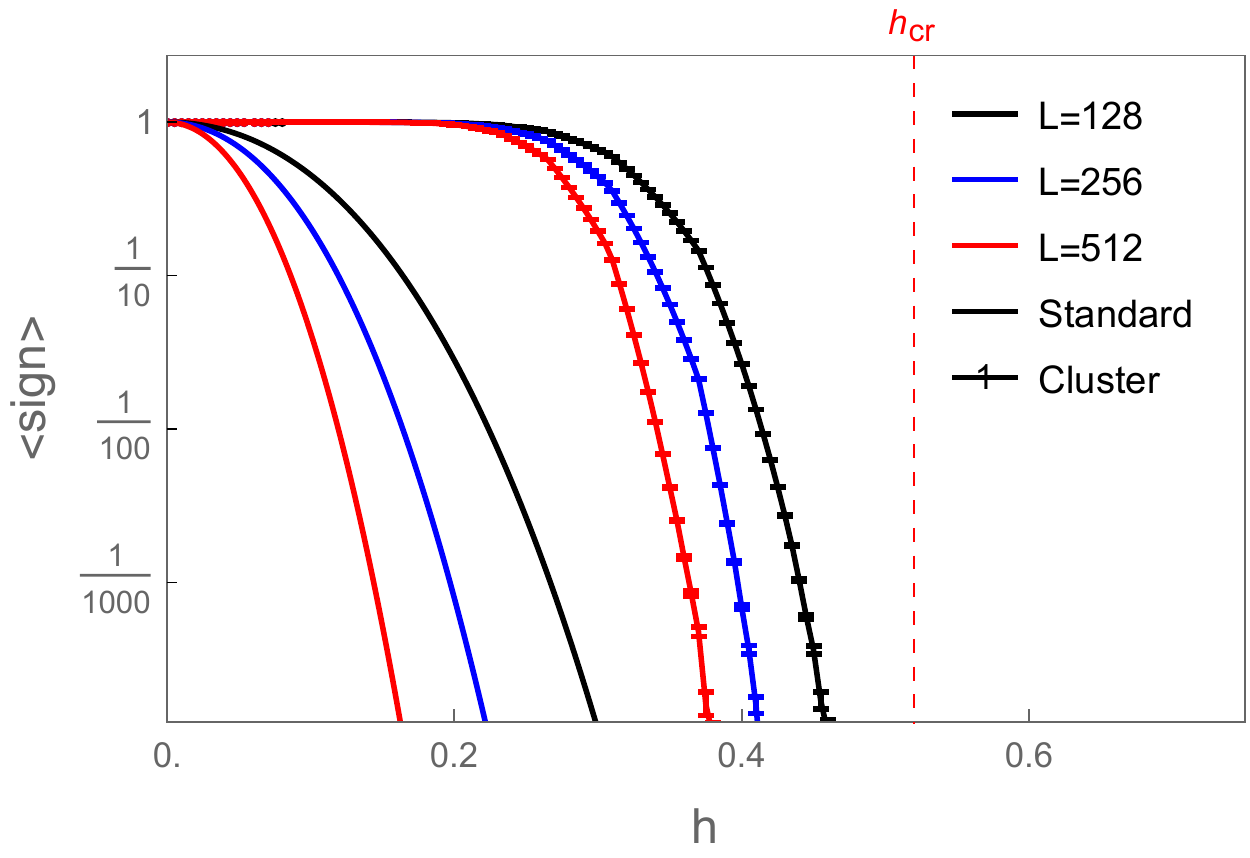}
\end{minipage}
\caption{The figure shows for different representations (spin, cluster and flux-variable representation) of the partition function \eqref{eq:partf1d} with $h'=0$ $h=\abs{h}\e^{\ii\,\pi/N}$ and for $N=2$ (left) and $N=3$ (right), the behavior of the average sign as function of $\abs{h}$, when approaching the first edge-singularity. As can be seen, the cluster representation allows one to get much closer to the edge-singularity before the average sign starts to drop dramatically.}
\label{fig:signproblemdiffrep1d}
\end{figure}

In \cite{Akerlund:2016myr} and \cite{Alford:2001ug}, it had already been shown that the flux-variable and cluster representations for the Potts model solve the sign-problem in \eqref{eq:pottspartf0} for certain values of $h$ and $h'$. Situations where the sign-problem is irreducible because the partition function becomes zero, were however not considered so far. In Fig.~\ref{fig:signproblemdiffrep1d} we show how the average sign in the spin-, flux- (for $N=2$) and cluster-representation of the partition function \eqref{eq:partf1d} for $h'=0$ and $h=\abs{h}\e^{\ii\,\pi/N}$ behaves as a function of $\abs{h}$ when approaching the value $h_{cr}$ where $Z_{N,L}\of{\beta,h\e^{\ii\,\pi/N},0}$ has its first zero. As can be seen, in the flux-variable representation the average sign drops almost as fast as in the ordinary spin-representation, whereas with the cluster representation, one can go much closer to the edge-singularity at $h_{cr}$ before the average sign starts to decrease dramatically.

\subsection{Oscillating/non-monotonic two-point functions in 2D}\label{ssec:oscillatingtwopointfunc}
\begin{figure}[h]
\centering
\begin{minipage}[t]{0.45\linewidth}
\centering
\begin{tikzpicture}
\node[inner sep=0pt,above right,opacity=1]{\includegraphics[width=\textwidth]{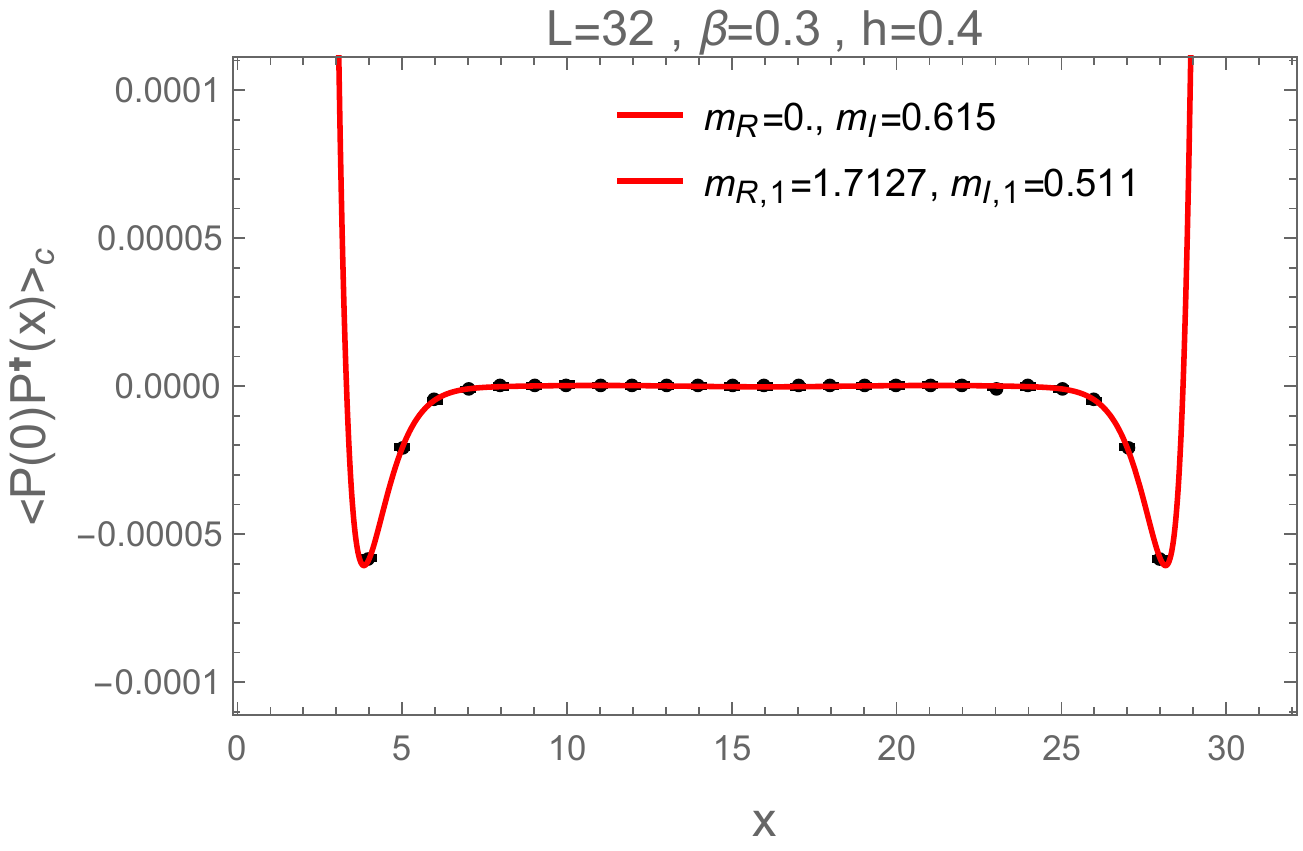}};
\end{tikzpicture}
\end{minipage}\hfill
\begin{minipage}[t]{0.45\linewidth}
\centering
\begin{tikzpicture}
\node[inner sep=0pt,above right,opacity=1]{\includegraphics[width=\textwidth]{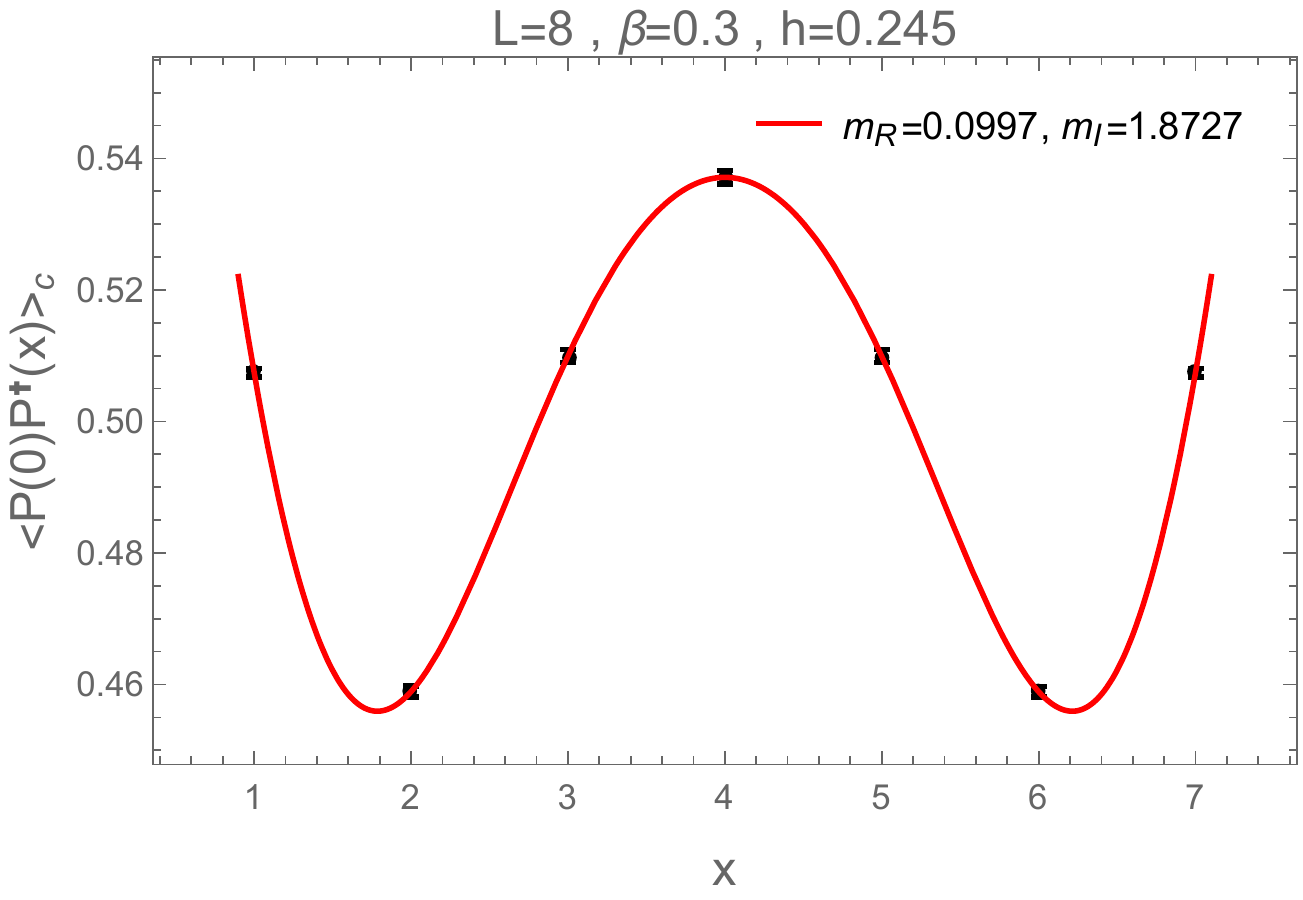}};
\end{tikzpicture}
\end{minipage}
\caption{The figures show for a $3$-state Potts systems in 2d, exemplary measurements of the two-point correlator \eqref{eq:twopointfunc} in the liquid (left) and crystalline (right) phase. In the left-hand figure, the red line corresponds to a two-mass fit, which yields the indicated two complex masses. In the right-hand figure, the red line corresponds to a single-mass fit.}
\label{fig:twopointcorr2d}
\end{figure}

Using the cluster representation \eqref{eq:clusterpartfgen} with $N=3$, $h'=0$ and $h\in\mathbb{C}$, we checked that the three prototype phases discussed in Sec.~\ref{ssec:edgesinganddolines} for the one-dimensional system, can also be found in two dimensions: Fig.~\ref{fig:twopointcorr2d} shows exemplary measurements of the two-point function \eqref{eq:twopointfunc} for $\beta=0.3$ and $h=0.4$ (left) in an system of size $V=32^2$, and for $\beta=0.3$ and $h=h\cdot\e^{\ii\,\pi/N}$ (right) in a system of size $V=8^2$, which correspond to the liquid and crystalline phases, respectively.

\section{Conclusion}\label{sec:conclusion}
We reviewed different ways to couple the $N$-state Potts model to a complex external field. This introduces in general a sign-problem, which in some case can be overcome again by changing to the flux-variable or cluster representation of the partition function, but if the partition function develops zeros for some values of the external field, then the sign-problem is \emph{irreducible} at these points. Although the sign-problem has at these points to be present in any representation (flux-var.,cluster,...), the rate at which the average sign drops, when approaching a zero of the partition function, can be significantly different for different representations: compared to the spin and flux-variable representations, one can, using the cluster-representation, get much closer to the zero of the partition function before the average sign starts to significantly deviate from unity and scales to zero.\\
We also discussed the relation between the sign problem and non-monotonic/oscillatory behavior of correlation functions.

\section{Acknowledgment}
We acknowledge grants of computer capacity from the Finnish Grid and Cloud Infrastructure (persistent identifier urn:nbn:fi:research-infras-2016072533).

\bibliography{lattice2017}

\end{document}